# Intelligent and Miniaturized Neural Interfaces: An Emerging Era in Neurotechnology

Mahsa Shoaran[1], Uisub Shin[1,2], MohammadAli Shaeri[1]

[1]EPFL, Lausanne, Switzerland, [2]Cornell University, Ithaca, NY

## Introduction

Over the past decade, there has been a growing interest in the development of intelligent neural interface systems-on-chip (SoCs) for a range of neurological disorders and emerging brain-machine interface (BMI) applications. The shift toward creating intelligent systems featuring on-implant signal processing, neural biomarker extraction, and AI has replaced prior efforts that primarily focused on raw neural signal acquisition and data compression for off-body processing [1-4]. Integrating complex functions into miniaturized neural devices presents significant opportunities for various applications, including therapeutic devices for central nervous system (CNS) disorders, peripheral nerve prostheses, spinal cord interfaces, and beyond. In this paper, we review the latest advancements in the development of CMOS-based integrated circuits (ICs) for three categories of intelligent neural prostheses, all featuring embedded signal processing on the implantable or wearable device. These categories include: 1) Neural interfaces for closed-loop symptom tracking and responsive stimulation; 2) Neural interfaces for emerging network-related conditions, such as psychiatric and memory disorders; and 3) Intelligent BMI SoCs for movement and communication recovery following paralysis. These developments mark the beginning of a dynamic field, and we anticipate the emergence of an even wider array of smart neural prostheses in the years ahead.

## Challenges toward future intelligent neural interfaces

Integration of advanced signal processing and machine learning (ML) algorithms on neural interface systems can significantly enhance the therapeutic potential of these devices in the future. For instance, AI-embedded neural interface technology has demonstrated its potential in enabling accurate, personalized symptom detection for patients with brain disorders, particularly epilepsy. More than a decade of active innovation in the development of ICs and AI algorithms has led to the creation of advanced systems, achieving greater than 95% sensitivity and specificity in epileptic seizure detection using hardware-efficient invasive or non-invasive SoCs [5-10]. Similarly, embedded neural biomarkers can guide the delivery of stimulation in a variety of neurological indications as they can represent the dynamic state of neuronal activity over time [11-14]. Furthermore, software-based AI algorithms have enabled increasingly complex BMI systems for rapid movement and communication recovery [15-18], with miniaturized hardware implementations recently emerging [8, 19]. While this progress is promising, there are still several challenges that must be addressed for the next generation of intelligent neural interface SoCs.

**Scalability:** Remarkable seizure detection performance has been achieved with hardware systems utilizing a limited number of sensing channels (8-24) from well-established EEG datasets, such as the pediatric CHB-MIT dataset [20]. A number of recent works expanded this to larger, intracranial EEG (iEEG)-based datasets with higher number of channels (≤128) and more difficult seizure patterns from adults with intractable epilepsy [21, 22]. The limited spatial resolution of electrodes in seizure detection or other symptom tracking systems may present practical challenges in real-world deployments due to the difficulty of gathering adequate neural activity relevant to pathological brain states. Moreover, accurate decoding of complex BMI tasks requires high-resolution intracortical or ECoG datasets. For instance, restoring dexterous (i.e., with a high degree of freedom) [23] or fine movements (i.e., with a high number of classes) like handwriting [16] or speech ability [17, 18, 24] necessitates the recording of data from hundreds of electrodes or even more.

**Efficiency:** The primary challenge in realizing a high-channel-count neural interface SoC lies in the increasing area and energy consumption of the hardware. The demand for compact, low-power chips becomes even more critical in the context of brain implants due to their invasive nature, physical placement, and the potential risk of heat generation, which could lead to tissue damage. Thus, ensuring that the employed AI algorithms are scalable is essential to process high-density neural data efficiently.

**Flexibility and adaptability:** Given the constrained area and power budget, current SoCs often rely on low-complexity AI models and small feature sets designed specifically for seizure detection. Embedding a flexible AI model along with an extensive set of features could enable the identification of more intricate patterns in brain activity and extend adaptability to various disorders beyond epilepsy. In addition, the characteristics of neural signals vary over time due to factors such as electrode movement, noise, and the dynamic nature of neural activity. In the context of epilepsy, there is also a need for algorithms capable of generalizing effectively across patients with limited seizure data, diverse seizure types, and datasets with high variability among patients. Consequently, the field has witnessed the emergence of novel techniques, including few-shot [25], one-shot [7], and zero-shot learning [26], as well as online learning algorithms and hardware [7, 10].

## 1. Closed-loop stimulation with real-time symptom tracking

The most prevalent application of AI in neural SoCs pertains to closed-loop stimulation systems, wherein stimulation parameters adapt in response to dynamic changes in the state of brain networks. In this context, the closed-loop NeuralTree SoC [8, 27] presented potential solutions to address several of the aforementioned challenges. A modular 256-ch front-end is implemented in the mixed-signal domain to enable area-efficient high-density neural sensing for effective AI model training, as shown in Fig. 1(a). Furthermore, as depicted in Fig. 1(b), a dynamic channel-selective inference scheme is implemented, in which only informative channels and features pertinent to specific disease states (identified via full-array high-density training) undergo selective processing. This approach helps to reduce hardware resource utilization during inference while preserving high accuracy. To enhance the SoC's versatility across multiple applications, a diverse set of biomarkers was integrated using hardware-friendly feature approximations and extracted in a disease-specific manner. A tree-structured neural network classifier, NeuralTree, employs hardware-efficient techniques such as network pruning and weight quantization. It is also trained with an energy-aware learning algorithm that penalizes power-demanding features, thereby further enhancing energy efficiency during inference. System-level co-design and optimization of circuits and algorithms resulted in significant improvements in channel count, compactness, and energy efficiency. The SoC also integrates a 16-ch, area-efficient high-voltage-compliant stimulator. In addition to seizure detection, the SoC demonstrated on-chip detection of tremors in

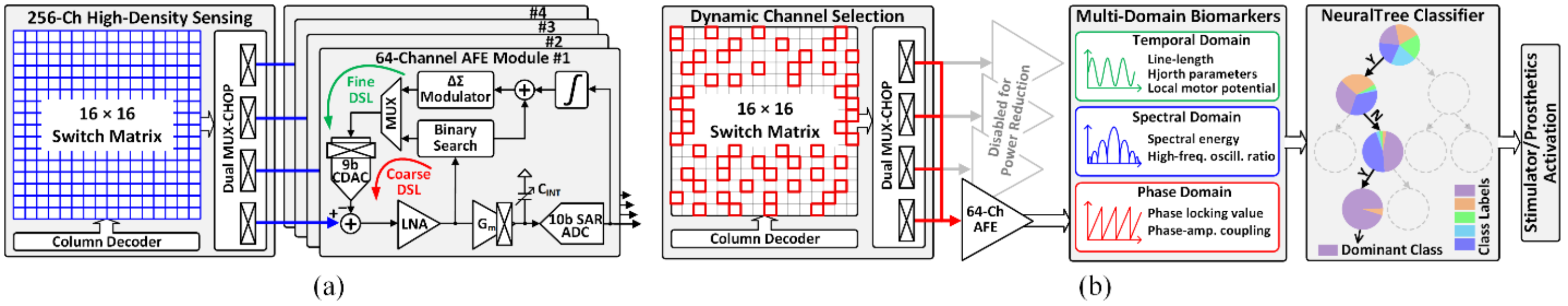

Fig. 1. The NeuralTree SoC [8]: (a) A modular 256-ch front-end enables high-density sensing for AI training. (b) Brain-state inference is executed along a single path of the tree, wherein the neural network in each node utilizes up to 64 dynamically selected features.

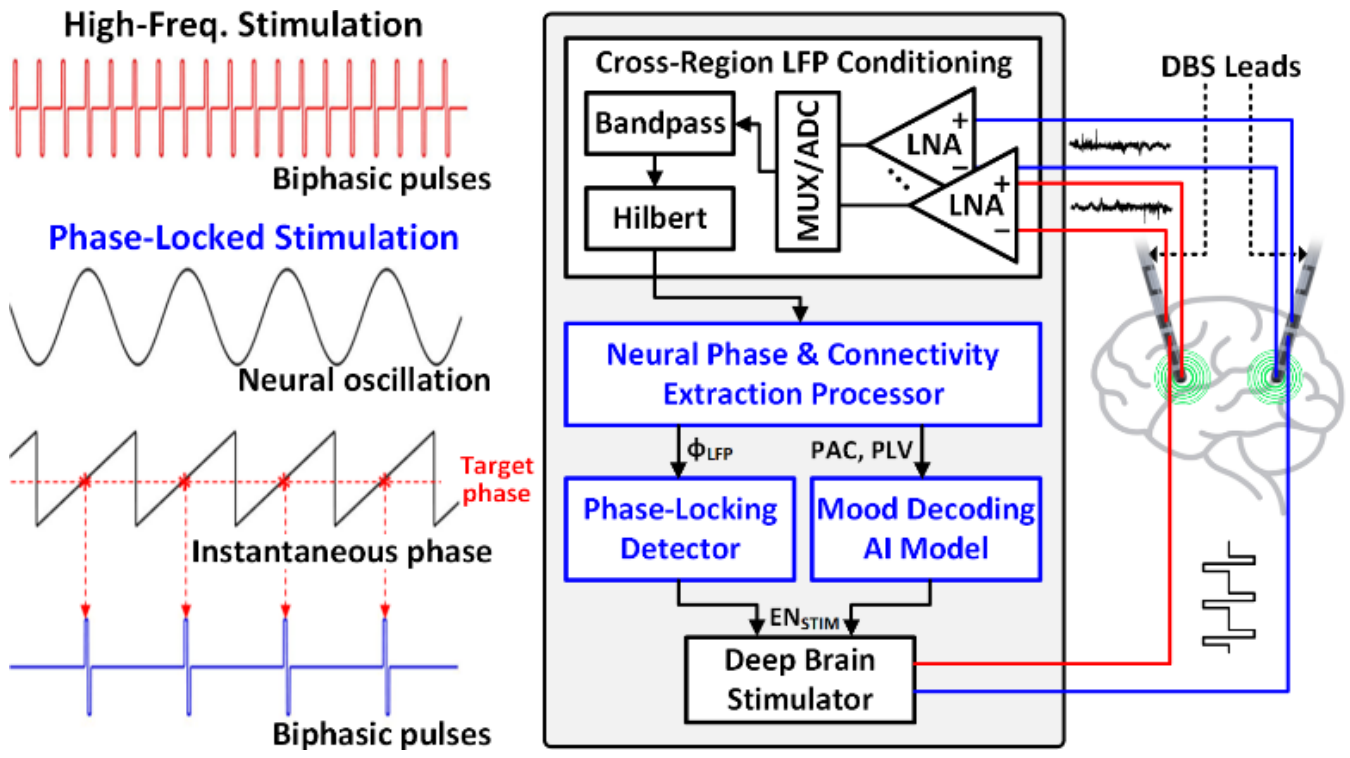

Fig. 2. Illustration of phase-locked DBS for regulating cross-region neural connectivity [43], and a potential advancement with on-chip AI for emerging applications such as mental disorders.

Parkinson's disease (PD) for the first time (Fig. 3, bottom-left), while the multi-class nature of the probabilistic NeuralTree [22] enables the potential deployment of this technology in prosthetic BMIs.

**Area-efficient SoC design**

Scaling the channel count up to several thousands requires a substantial reduction in the die area occupied by neural recording amplifiers. To address this challenge, traditional AC-coupled neural amplifiers [28] have given way to mixed-signal or digitally intensive designs, which scale more gracefully with advanced CMOS technology nodes. The 0.025mm$^2$/ch front-end in [4] introduced a mixed-signal electrode DC offset cancellation method, effectively replacing an area-consuming analog integrator. The adoption of an open-loop architecture for signal amplification facilitated a significant reduction in the input capacitors, at the expense of increased gain mismatch. Recently, hardware sharing through time-division multiplexing (TDM) has become a popular design choice for high-density area-efficient neural recording. For instance, the NeuralTree SoC's 256-ch 0.004mm$^2$/ch front-end utilized four 64-ch TDM modules to perform high-density neural sensing for AI training. A two-step fast-settling DC servo loop (DSL) rapidly canceled DC offsets from multiplexed inputs, which were dynamically reconfigured for each 1s window of channel-selective inference. Another 256-ch ECoG recording front-end adopted multiplexing and reported a compact area of 0.001mm$^2$/ch (with an off-chip multiplexed switch matrix) [29]. A hardware-efficient DSL coarsely canceled electrode DC offsets, while neural signals with residual offsets were digitized by incremental delta-sigma ADCs. In time-shared architectures, the requirement for a wider bandwidth amplifier introduces elevated noise folding from electrodes. Therefore, the choice of an optimal degree of multiplexing is a critical decision that requires careful consideration of various factors. These include electrode impedance, available silicon area, desired signal modality and fidelity, along with the noise resilience inherent in AI models.

**Other algorithm and hardware considerations**

As the channel count increases, the associated overhead for AI training can become prohibitively high. For instance, training the classifier on large neural data offline could demand a significant amount of energy consumption for raw data transmission. This may require the reemergence of lightweight compression techniques (e.g., compressive sensing [3]) or replacing raw data with feature transmission to lower the telemetry power during initial model training. The integration of advanced security measures such as data encryption will become imperative for such devices to protect sensitive data and preserve privacy. Moreover, the classification performance may degrade over time as the pattern of pathological brain activity changes. Recent SoCs have introduced several approaches to reduce retraining overhead and maintain long-term accuracy. The seizure detector in [10] employed unsupervised learning to retrain a logistic regression classifier online. While this approach performed well on the CHB-MIT dataset, the basic linear model employed may not be optimal for handling more complex seizure patterns. Another interesting concept was introduced in [26], where the SoC achieved patient-independent seizure detection by training a convolutional neural network on pre-existing datasets. Fine-tuning of model parameters was subsequently performed online to account for inter-patient variability in seizure patterns.

**2. New paradigms for closing the loop**

Deep brain stimulation (DBS) stands as a well-established therapeutic approach for movement disorders. Continuous high-frequency DBS (e.g., 130Hz) has proven effective in suppressing motor symptoms associated with PD and essential tremor. Furthermore, DBS has exhibited potential in a growing range of applications in recent years, spanning from movement disorders to epilepsy, stroke, psychiatric and memory-related conditions [30]. However, open-loop DBS may result in various side effects, excessive energy consumption and battery usage, and reduced effectiveness over time [31, 32]. Recent evidence suggests that a novel approach utilizing closed-loop, phase-locked DBS [33, 34] can be as effective in addressing movement disorders [35] and holds potential for treating psychiatric conditions like major depression [36]. While the biomarker-driven closed-loop approach (e.g., NeuralTree) is an effective solution for disorders with well-established biomarkers such as epilepsy and PD, it may not be equally suitable for psychiatric and memory disorders with more intricate underlying mechanisms. Alternatively, phase-locked DBS delivers bursts of stimulation precisely locked to specific phases of neuronal oscillations, as illustrated in Fig. 2 (left). This approach holds the potential to enhance therapeutic efficacy while minimizing side effects and enabling prolonged effectiveness, thanks to enhanced plasticity [37].

Furthermore, brain connectivity, whether within- or cross-region, plays a pivotal role in detecting pathological brain states across various neurological and psychiatric disorders, particularly those that impact distributed brain networks. For instance, it was shown that epileptic seizures manifest spatial and temporal changes in cross-channel phase synchronization [38]. Also, excessive phase-amplitude coupling (PAC) has been observed in patients with PD [39]. Moreover, conditions like depression [40], post-traumatic stress disorder (PTSD) [41], and Alzheimer's disease [42] have shown disruptions in network connectivity, emphasizing the significance of connectivity analysis in understanding and diagnosing these complex disorders.

Inspired by these neuroscientific findings, a recent closed-loop neuromodulation SoC [14, 43] introduced novel stimulation paradigms, wherein neural connectivity within or across regions is monitored continuously and regulated through phase-locked DBS. While earlier seizure detection SoCs [44, 45] demonstrated on-chip computation of connectivity metrics such as PAC and phase-locking value (PLV) using iterative vector processing with CORDICs, their high accuracy comes at the cost of excessive power consumption (>200μW). In [14, 43], the complex nonlinear functions for phase and amplitude computation were efficiently approximated using a last-bit accurate linear arctangent algorithm and the $\ell_\infty$-norm, resulting in >60% power savings for PAC/PLV extraction without compromising accuracy. This SoC demonstrated the first-in-literature phase-locked neurostimulator and was validated in-vivo in rats within regions associated with fear and anxiety (Fig. 3, bottom-right). In addition, a multi-mode stimulation control is supported through various combinations of phase-locking events and thresholded connectivity measures, which may be useful for treating different neurological conditions in the future.

Further advancements in this technology can be achieved by integrating an AI model and utilizing its decision to guide phase-locked DBS control, as envisioned for the treatment of mental disorders in Fig. 2 (right). Depending on the target application, it may be necessary to explore new types of biomarkers and connectivity measures to improve detection accuracy and therapeutic efficacy. In addition to traditional classifiers with handcrafted features, deep learning models that automatically mine features from raw neural activity may offer promising alternatives. System-level innovation across ICs and learning algorithms would be critical to efficiently integrate such complex models into resource-constrained brain implants. Lastly, there has been a continued demand and ongoing efforts to realize concurrent brain sensing and stimulation capabilities for uninterrupted brain-state detection in closed-loop settings. This poses numerous circuit/algorithm design challenges, including high-dynamic-range neural recording, fast artifact recovery, and stimulation artifact cancellation—all of which must be achieved without consuming excessive area and power.

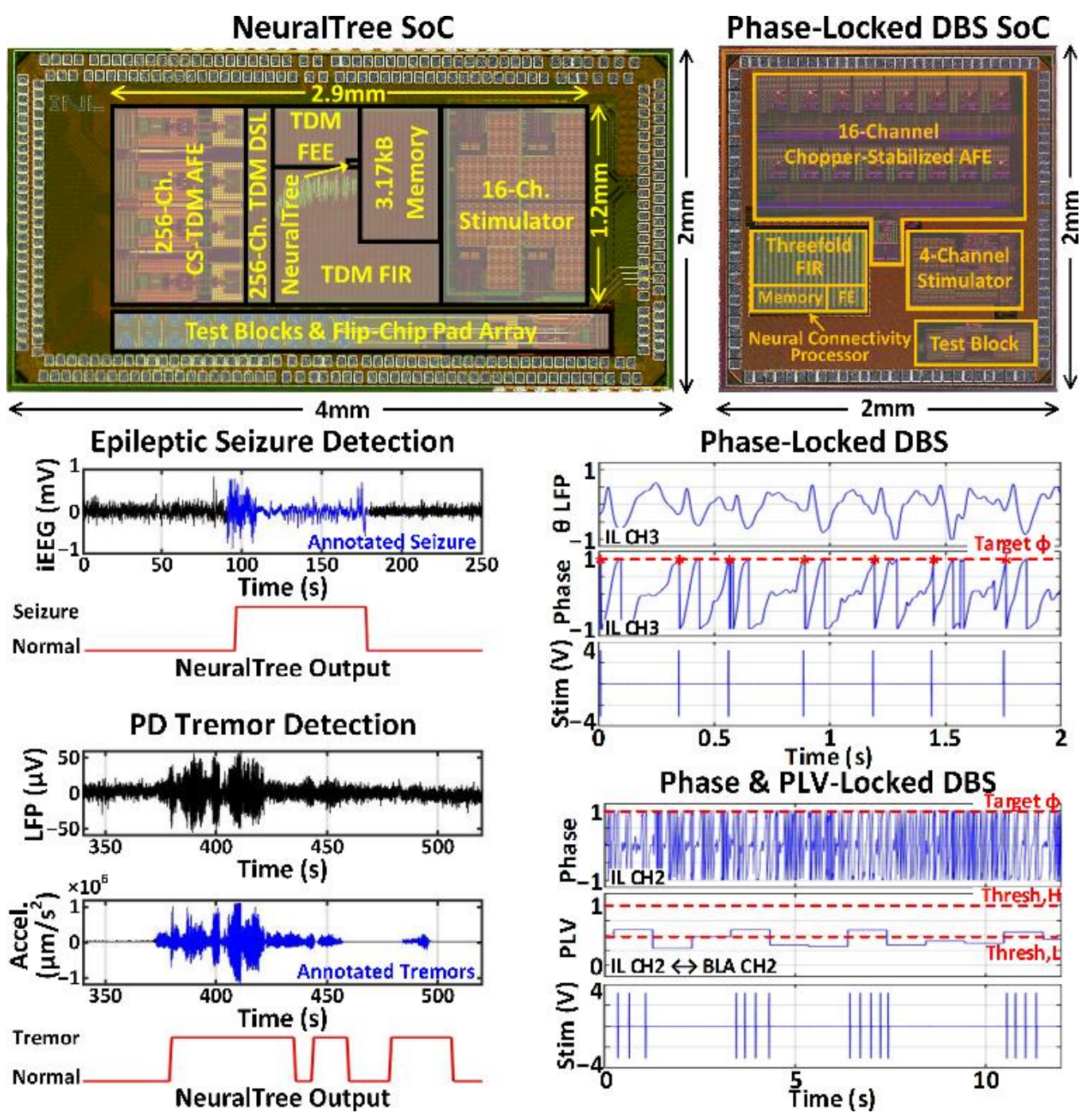

Fig. 3. Chip micrographs and experimental results of the NeuralTree and phase-locked DBS SoCs [8, 43].

## 3. Neural signal processing SoCs for BMI applications

In another context, intelligent neural interfaces have the potential to significantly improve prosthetic BMIs, with the goal of restoring lost motor or communication abilities for paralyzed patients. As efficiency and miniaturization continue to improve, these interfaces can play a pivotal role in facilitating the translation of BMIs into daily lives of patients beyond clinical settings. Moreover, recent advancements in ultra-high-density microelectrode arrays, dense cortical grids, and flexible polymer threads (e.g., Neuropixels probe and Neuralink systems) are enabling unprecedented levels of sensing resolution in neural interfaces [27, 46-48]. While this has the potential to revolutionize the functionality of BMIs, transmitting such a large amount of data for offline processing and analysis comes with significant power consumption and security risks. In realizing fully implantable, energy-efficient BMIs, it is crucial to implement on-chip signal processing and/or data compression algorithms with minimal power and area overhead, while preserving essential information. To this end, a variety of on-chip signal processing approaches have been introduced for BMIs, ranging from spike detection [49-52] and sorting [53-60] to feature extraction [8, 49, 56, 57] and movement decoding [8, 19, 61-63].

Importantly, next-generation intelligent BMIs must incorporate on-chip decoding capabilities to support a variety of applications. Yet, the challenge arises from the fact that current software-based BMIs often utilize complex models like recurrent neural networks (RNNs) with extensive parameter counts. Efficient integration of such models in resource-limited implantable or wearable platforms presents a significant challenge. This model complexity stems from the distributed nature of neural activity associated with motor functions, resulting in significantly higher data dimensionality compared to more typical problems like spike sorting. In addition, the complexity of BMI tasks (as indicated by factors such as the number of classes or degree of freedom) demands the use of advanced AI models rather than relying solely on basic signal processing and data clustering techniques. Thus, a primary hurdle in achieving fully-implantable, miniaturized, and low-power BMIs is the development of neural decoders capable of effectively learning from intricate high-dimensional neural data and seamlessly integrating them with brain implants.

### Spike detection and sorting SoCs

As an initial step, the detection of spiking activity is crucial to analyze intracortical data at the level of individual neurons. A common approach is to detect threshold-crossing events either in the time or nonlinear energy operator (NEO) transform domains as spikes [49-52, 55, 59, 60, 64]. NEO emphasizes sharp, high-energy signal changes, leading to superior detection accuracy and reduced sensitivity to thresholds, making it a popular choice for spike detection chips. While the threshold level can be computed during off-chip training [60], there is a growing interest in on-chip unsupervised approaches that adaptively compute the threshold [50, 52, 55, 59]. Traditionally, the threshold level can be adjusted online using statistical measures such as the mean, median, or standard deviation of the averaged signal over a sliding time window [51, 59, 64, 65]. To improve detection accuracy, a hardware-efficient unsupervised dual detector was introduced in [50]. Utilizing two detection pathways to discern signals in both high- and low-noise scenarios, this approach achieved a record spike detection accuracy of 97.4% on the standard Wave_Clus (Quiroga) dataset. Assuming a known spiking rate for each specific brain region, a threshold was efficiently calculated in [52], leading to reduced hardware costs albeit with a trade-off in accuracy. While most spike detection SoCs have been evaluated on the single-channel, synthetic Wave_Clus dataset, validating their performance on emerging high-density datasets (e.g., those recorded by Utah arrays or Neuropixels probes) is crucial to demonstrate their robustness and reliability for future clinical and prosthetic applications.

Standard AI tools play a significant role in spike sorting, the process of assigning individual spikes into distinct clusters based on waveform similarities. To achieve this, various features are extracted from the detected spikes and subsequently classified into separate neuronal classes. For instance, static time-domain features of spike waveforms such as spike peaks and their derivatives can be extracted to reduce data dimensionality and the hardware complexity of classifiers [64, 66]. A salient feature selection method was proposed in [56] to dynamically select a minimal subset of spike features based on the highest class discrimination. Alternatively, an $\ell_2$-normalized convolutional autoencoder with a fully connected layer aimed to identify informative features for the clustering process [57]. The extraction of informative features not only enhances hardware efficiency but also improves accuracy and robustness in the spike sorting process.

To classify spike waveforms, several hardware-efficient methods such as oblique decision trees [66] and window discrimination [56] have been employed in literature. While these designs are not self-sufficient and require off-chip training, a current technology direction is to employ unsupervised spike sorters to eliminate the need for off-chip training. K-means and its variants are widely used for this purpose owing to the simplicity of the model. These methods assign each data point to a cluster by evaluating the minimum distance, utilizing various metrics such as $\ell_1$-norm [55] and $\ell_2$-norm distance [59], cosine similarity [57], or correlation coefficient [58]. Subsequently, the spike sorter updates the spike cluster using the newly acquired data. A recent spike sorter was developed specifically for the 384-ch Neuropixels probe with closely-spaced electrodes [59]. Given that each neuronal spike may potentially be recorded by multiple neighboring channels, this sorter selects the channel with the highest amplitude while discarding redundant spikes captured by other channels. This approach has shown outstanding accuracy of 97.7% on a pre-recorded Neuropixels dataset. Recent spike sorting SoCs have achieved power levels in the order of 1-3μW/ch and 0.001-0.02mm$^2$/ch silicon area [59, 60].

In the absence of ground truth datasets in spike sorting, current evaluations rely on synthetic datasets or real recordings with manually created labels, which hinders reliable accuracy comparisons across systems. Additionally, challenges such as the separation of temporally overlapping spikes sensed by a single electrode and low performance in high-noise scenarios are areas that must be addressed in the future.

Although many interesting spike detection and sorting approaches have been proposed and implemented at the chip level, a significant hurdle is on-chip decoding of neural activity. While spike sorting deals with relatively simple tasks, involving low-dimensional data and a limited number of classes, neural decoding necessitates the analysis of ultra-high-dimensional data with significantly greater task complexity—a challenge yet to be fully addressed.

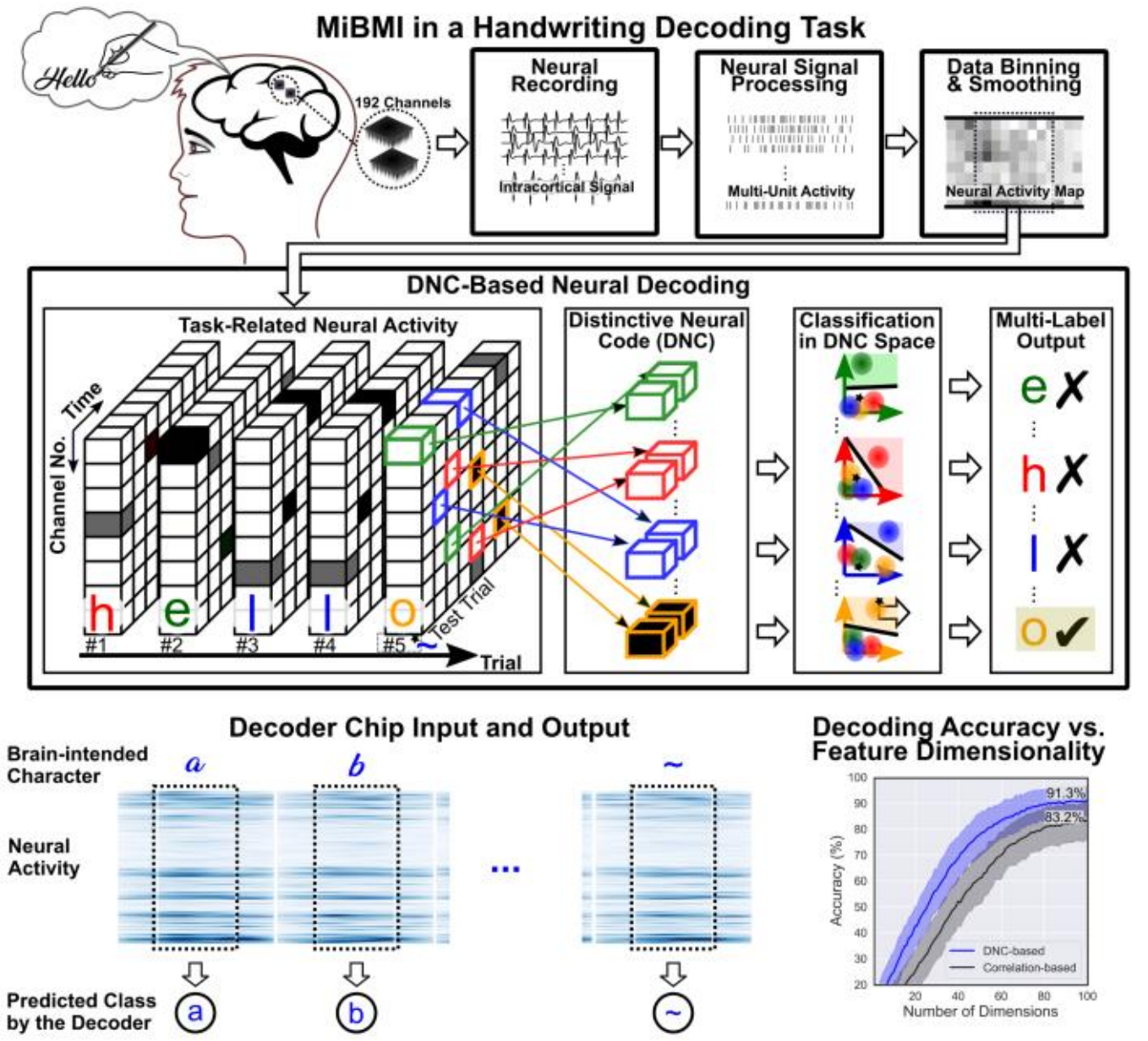

Fig. 4. The concept of distinctive neural codes (DNCs) and the DNC-based decoding results on the 31-class handwriting task [19].

### Neural SoCs with embedded movement decoding

Recent innovations in BMI technology have shown a remarkable potential for transforming the lives of individuals with paralysis, particularly those who have lost the ability to move or communicate. Empowered by advanced AI algorithms, recent BMIs have demonstrated the decoding of brain activity associated with various movements and actions, including gait, reach-and-grasp, cursor control, typing, handwriting, and speech [15-18, 23, 24, 67-70]. Moreover, notable breakthroughs have showcased the potential of BMIs in studying cognitive processes such as decision-making and neural plasticity [69, 70], providing insights into the complexities of cognitive functions. However, these BMIs depend on powerful yet bulky computers with limited mobility, making them impractical for everyday use and daily life settings of patients.

Only a handful of papers have reported on-chip decoding for BMI applications. In [61], a neuromorphic SoC was developed to decode four-class neural activity evoked by cortical stimulation to control a robotic arm, consuming a substantial chip area and power. Alternatively, [62] demonstrated a 128-ch extreme learning machine (ELM) for decoding finger movements. The hidden layer (i.e., a random projection layer) was implemented on chip, while the processing of the output layer to generate decisions was conducted off-chip. The 93-ch intracortical BMI system in [63] decoded finger movement intentions using spiking band power (SBP) features within the 0.3–1kHz frequency range and a predictor based on a steady-state Kalman filter (SSKF). While this system achieved high accuracy in closed-loop finger movements, it exhibited a notable latency of up to 2.4s and utilized a commercial recording system (Intan). The high-density NeuralTree SoC [8] enables ECoG-based finger movement classification, but it lacks high-bandwidth spike recording capability, which is crucial for more complex motor decoding tasks.

Although the mentioned BMIs have shown high accuracy in decoding basic movements, there is an increasing need for the development of advanced on-chip decoders, possibly using interpretable models, to handle more intricate BMI tasks like handwriting and speech. Utilizing interpretable models is beneficial not only for gaining insights into brain functions but also for advancing prosthetics development by establishing solid scientific foundations. Moreover, these models often yield meaningful results, facilitating thorough validation of decoder predictions. Additionally, in the development of implantable BMIs, the integration of a low-power, custom-designed neural recording unit alongside the on-chip decoder can substantially reduce power consumption and device form factor.

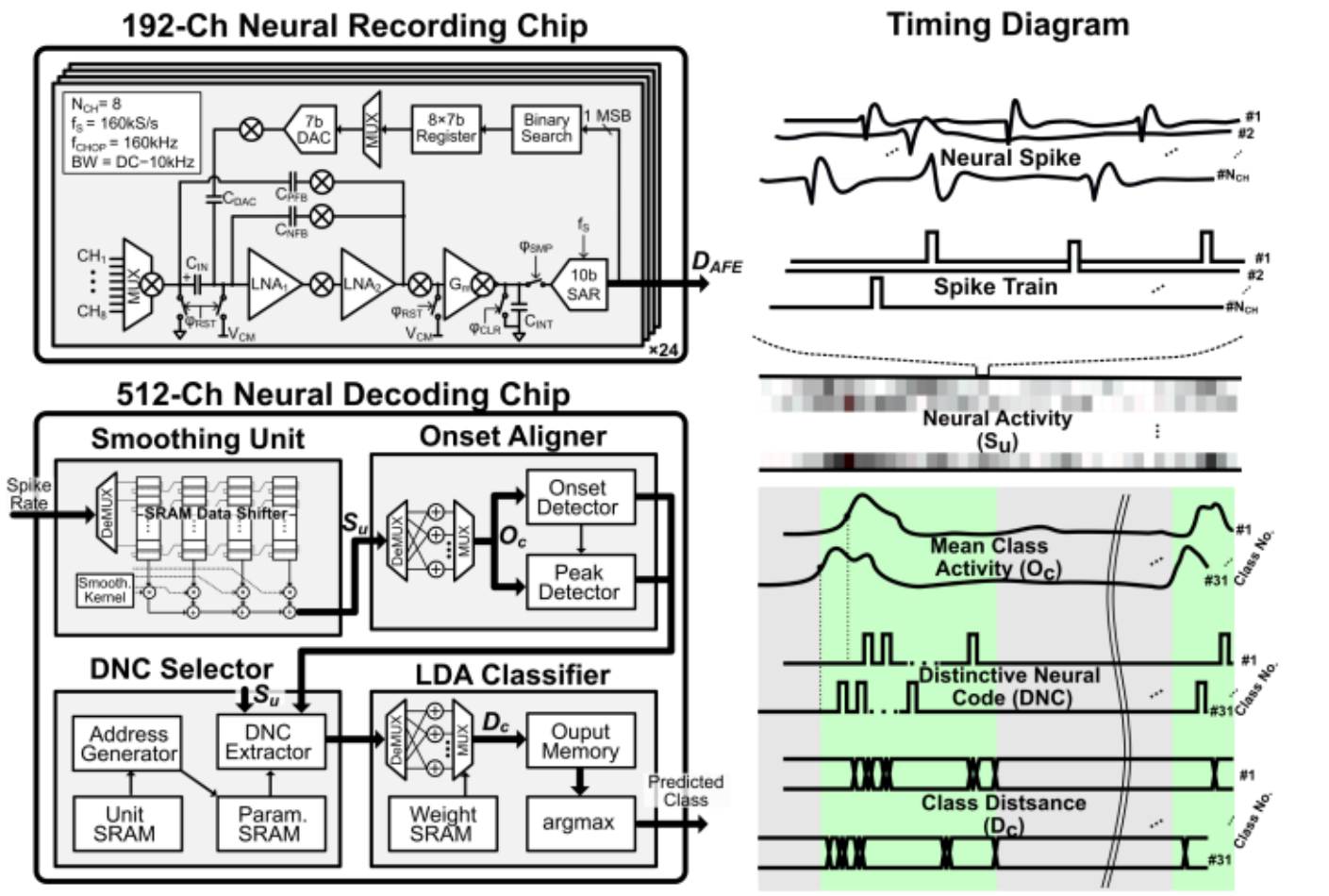

Fig. 5. The hardware architecture of the MiBMI chipset and related timing diagram [19]. The integrated 192-ch front-end chip performs area-efficient neural recording. Upon smoothing, detecting activity onset, and alignment, the decoder extracts DNCs and classifies the attempted character.

### A miniaturized brain-to-text BMI

A new generation of BMIs [16-18, 24] strives to greatly improve the restoration of lost communication abilities such as writing and speech for paralyzed patients. Thanks to the fine motor skills required for tasks like handwriting, decoding such complex movements can be achieved at considerably higher speeds compared to conventional BMIs that primarily predict simple point-to-point movements. However, these tasks often demand sophisticated AI models to decode neural activities linked to dexterous motor skills, which poses challenges for efficient integration within brain implants. To address this challenge, a recent miniaturized BMI chipset (MiBMI) [19] employed the new concept of Distinctive Neural Code (DNC) as a promising solution to decode attempted handwriting. Figure 4 illustrates the concept of a decoder utilizing DNCs in the context of a conceptual 31-character handwriting task [16]. Inspired by the brain's saliency model of attention, the most distinctive features of neural activity (i.e., DNCs) are selected, effectively transforming a high-dimensional state space to a lower-dimensional subspace. This allows for the accurate differentiation of various classes in complex BMI tasks, while eliminating the need for resource-intensive pre-processing steps like time warping [16]. Another significant advantage is that, due to the rich information content of DNCs, complex multi-class decoding tasks can be achieved using simple classifiers (e.g., linear discriminant analysis or LDA) with 1750× fewer parameters compared to the RNN used in [16].

Figure 5 shows the block diagram of the brain-to-text decoder exploiting DNCs. It receives a 512-ch neural input (i.e., spike counts), which is then smoothed to mitigate the decoder's sensitivity to misalignments and noise (Fig. 5, right). A subset of neural activities is selected following a movement attempt, and their mean class activity is computed for onset detection and alignment. This is followed by real-time extraction of DNCs for each class. This method reduces data dimensionality by 51200× at the decoder's output, while also lowering the training time of the decoder. Moreover, through DNC extraction and memory sharing, this approach can drastically reduce computational requirements, performing 320× fewer MAC computations and utilizing 100× less memory compared to a traditional LDA model. The fabricated decoder occupied a compact area of 0.75mm$^2$, consumed 223μW, and achieved a notable accuracy of 90.8% in the 31-class motor decoding task.

To demonstrate the feasibility of a complete BMI chipset, a compact 192-ch neural recording chip was additionally integrated, as illustrated in Fig. 5 (top-left). Each of the 24 multiplexed modules

Table I. Design specifications and performance summary of the state-of-the-art spike sorting and BMI SoCs.

| Parameter | Front. Neurosci.'16 [61] | TBioCAS'16 [62] | TBioCAS'19 [55] | J. Neurosci. Methods'19 [54] | TBioCAS'22 [60] | JSSC'22 [8] | TBioCAS'22 [63] | JSSC'23 [59] | ISSCC'24 [19] |
|---|---|---|---|---|---|---|---|---|---|
| Process (nm) | 180 | 350 | 130 | 40 | 22 | 65 | 180 | 22 | 65 |
| Supply Voltage (V) | 1.8 | 0.6/1.2 | 1.09 | 1.1 | 0.5-0.8 | 1.2 | 0.625 | 0.59 | 1.2 |
| # of Channels | 15 | 128 | 10 | 16 | 16 | 256† | 93 | 384 | 192/512 |
| Features | - | - | Symmlet-2 Wavelet | - | Adaptive filtering | Spectral powers, LMP | SBP | Spike peak and derivatives | DNC |
| Decoding Model | SNN | ELM | K-means | BOTM | K-means | NeuralTree | SSKF | K-means | LDA |
| Decoding Task | Prosthetic Control | Finger Movement | Spike Sorting | Spike Sorting | Spike Sorting | Finger Movement | Finger Movement | Spike Sorting | Handwriting |
| Input Signal | Spike Train | Spike Train | Intracortical | Intracortical | Intracortical | ECoG | Intracortical | Intracortical | Spiking Rate |
| # of Classes | 4 | 12 | 5‡ | 4‡ | 8‡ | 6 | Continuous | 2‡ | 31 |
| Accuracy (%) | 50−70 | 99.3 | N/A | 93.4* | 94.12* | 73.3 | 100(1D), 96(2D) | 89.5*, 97.7+ | 90.8[A], 91.3[B] |
| AI Area/ch ($mm^2$) | 2.86$ | 0.133$^ | 0.08 | 0.0175 | 0.014 | 0.02$ | 0.097 | 0.0013 | 0.0015 |
| AI Power/ch (μW) | 267 | 0.01^# | 56.9 | 19 | 2.79 | 4.24 | 6.2(1D), 6.3(2D) | 1.78 | 0.44 |

† Selective 64-ch processing ‡ Maximum number of clusters/ch * Wave_Clus dataset + Neuropixels dataset $ Estimated active area
^ Excluding the off-chip output layer # Power measured with 40 input channels A Measured on 620s data B Simulated on 10.7-hr data records 8-ch broadband (10kHz) neural activity containing both action potentials (APs) and local field potentials (LFPs). Through binary search in the feedback, the DC offsets from multiplexed inputs are suppressed to a linear range of the amplifiers. In this time-shared architecture, inter-channel crosstalk is effectively reduced by periodically resetting the capacitors, while the resulting kT/C as well as flicker noise are up-modulated by chopping and subsequently filtered out. The 192-ch recording chip occupied an active area of 1.7mm$^2$ (0.009mm$^2$/ch) and consumed 660μW (3.44μW/ch). This demonstrates the potential for replacing power-demanding and bulky off-the-shelf recording units, enabling the realization of fully implantable and miniaturized BMIs.

Table I summarizes the state-of-the-art BMI SoCs with embedded AI. For spike sorters, the maximum number of spike clusters per channel is reported as a measure of task complexity. As the technology matures, there has been a clear shift toward the development of high-channel-count, low-power, and miniaturized BMIs. With advances in feature engineering, efficient data representation, and on-chip AI, the complexity of decoding tasks has also evolved—from single-digit-class spike sorting and finger movement to more intricate 31-class character decoding.

## New opportunities in AI-enhanced BMIs

Current BMIs primarily rely on powerful external computers (e.g., CPUs or GPUs) for neural signal processing. On the other hand, the next-generation BMIs aim to become unintrusive, low-power implantable devices, ideal for daily-life applications in patients with disabilities. These devices will seamlessly incorporate efficient AI models to learn and decode complex neural data, while enhancing patient privacy and reducing reliance on data telemetry.

To serve as practical neuroprostheses for patients, decoding models must offer both low latency and high accuracy. Closed-loop BMIs can further enable bidirectional communication between the neural interface and the user, allowing real-time adjustments and enhancing prosthetic device functionality. For instance, incorporating tactile and sensory feedback mechanisms can provide patients with a more natural and dexterous control experience over the BMI. Furthermore, hybrid BMIs capable of decoding multiple neural signal modalities hold the potential to enhance the robustness and decoding accuracy of these interfaces in the future.

## Conclusion

This paper reviewed the major challenges in advancing intelligent neural interface technology, including closed-loop neuromodulation and prosthetic BMIs, with a special focus on integrated circuits and on-chip AI. We delved into various aspects such as the development of high-resolution neural recording ICs, exploration of new biomarkers and closed-loop stimulation methods for emerging neurological applications, hardware realization of AI models, among others. We also discussed various IC and SoC design examples that aim to address these challenges either individually or collectively, in addition to providing an outlook on other important problems that remain to be tackled. Continued research and development efforts led by circuit designers and AI experts are anticipated to further advance this new era in neurotechnology, ultimately enhancing the quality of life for a diverse population of patients around the world.